\documentclass[preprint,aps,longbibliography,superscriptaddress]{revtex4-1}
\usepackage{graphicx,epsfig,subfigure,dcolumn,bm,mathrsfs,amsmath,amssymb,xcolor}

\makeatletter
\renewcommand{\p@subsection}{}
\renewcommand{\p@subsubsection}{}
\makeatother

\newcommand{\ud}{\,\mathrm{d}}
\newcommand{\change}[1]{\textcolor{black}{#1}}

\begin{document}

\title{Capillary Filling in Closed-end Nanotubes}

\author{Chen Zhao}
\affiliation{Center of Soft Matter Physics and Its Applications, Beihang University, Beijing 100191, China}
\affiliation{School of Physics and Nuclear Energy Engineering, Beihang University, Beijing 100191, China}

\author{Jiajia Zhou}
\email[]{jjzhou@buaa.edu.cn}
\affiliation{Center of Soft Matter Physics and Its Applications, Beihang University, Beijing 100191, China}
\affiliation{Key Laboratory of Bio-Inspired Smart Interfacial Science and Technology of Ministry of Education, School of Chemistry, Beihang University, Beijing 100191, China}

\author{Masao Doi}
\affiliation{Center of Soft Matter Physics and Its Applications, Beihang University, Beijing 100191, China}
\affiliation{Beijing Advanced Innovation Center for Biomedical Engineering, Beihang University, Beijing 100191, China}

\begin{abstract}
Capillary filling in small length scale is an important process in nanotechnology and microfabrication.
When one end of the tube or channel is sealed, it is important to consider the escape of the trapped gas.
We develop a dynamic model on capillary filling in closed-end tubes, based on the diffusion-convection equation and Henry's law of gas dissolution.
We systematically investigate the filling dynamics for various sets of parameters, and compare the results with a previous model which assumes a linear density profile of dissolved gas and neglect the convective term.
\end{abstract}

\maketitle





\section{Introduction}

Capillary filling is a common phenomenon in our daily life.
One can observe this phenomenon when inserting a glass tube in a water tank or dipping a dried paper tower into a fluid \cite{dBQ, Bico2003}.
In both cases, the fluid starts to penetrate into the tube/paper, increases the contact area with the wetted material, and reduces the interfacial energy.
Capillary filling also plays an important role in biology, and it is responsible for the phenomenon that water and nutrients are drawn from the soil by the plant root up to the stalks and leaves \cite{Jensen2016}.

Near one hundred years ago, the dynamics of capillary filling has been worked out by Lucas and Washburn \cite{Lucas1918, Washburn1921}.
The celebrated result of Lucas-Washburn equation predicts that in an open cylindrical tube, the filling length $h$ of a Newtonian fluid is proportional to the square root of time $t$,
\begin{equation}
  \label{eq:LW}
  h(t) = {\rm constant} \times \sqrt{t}.
\end{equation}
The theory was based on the macroscopic length scale.
Surprisingly, Lucas-Washburn dynamics of $\sqrt{t}$ dependence can be extended to nanometer scales \cite{Dimitrov2007a, Shin2007, Ouali2013, YaoYang2017}.
Capillary filling in small length scale is important because of potential applications in nanotechnology and microfabrication.
For example, nanoporous materials play an important role in applications of DNA translocation \cite{Meller2001Voltage,Rabin2005DNA}, nanofluidic transistors \cite{Karnik2005}, templates for nanoparticle self-assembly \cite{Alvine2006Solvent}, and sensors for chemical agents \cite{Novak2003Nerve}.

One outstanding problem of capillary filling in nanometer scale is that the front factor in Eq.~(\ref{eq:LW}) measured in experiments is consistently smaller than that calculated from the material properties \cite{Chauvet2012}.
Many studies had been carried out to elucidate this discrepancy.
For example, Tas et al. \cite{Tas2004Capillary} dealt with the issue by the electroviscous effect.
They excluded the effect of bubble formation by selecting isopropanol and ethanol as their experimental fluid, where no apparent bubble formation was observed.
Thamdrup et al. \cite{Thamdrup2007} observed that the formation of air bubbles further reduces the filling speed, and they introduced a correlation between the bubble density and the filling rate.
A comparison between analytic results and experimental results also showed that the electroviscous effect is not the only cause of the reduction in filling speed \cite{Phan2010,Phan2009}.

One common system involves fluid penetrating a tube or a channel with closed ends.
Phan et al. \cite{Phan2010} highlighted the importance of gas dissolution in the fluid.
They performed experiments with ethanol and isopropanol to investigate the capillary filling process of closed-end nanochannels.
They developed a model based on Henry's law and explicitly considered the gas dissolution and diffusion.
One assumption of the model is that the gas number density in the fluid is linear from the front to the rear end.
The agreement between the experiments and the model seems to be reasonable.

Here we present a more careful investigation of fluid filling in a cylindrical nanotube with one end is closed.
In our soluble gas model, we do not make the assumption of a linear profile.
We also include the convection term in our model and its influence on the dynamics is important for certain parameter sets.
The reminder of this article is organized as follows: In Section 2, we derive different models of capillary filling in details.
We then present numerical results of various models in Section 3.
Finally, we conclude in Section 4 with a brief summary.

\section{Theoretical Models}
\label{sec:theory}

\begin{figure}[htbp]
  \centering
  \includegraphics[width=0.6\columnwidth]{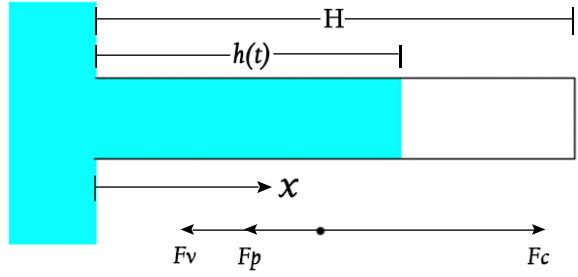}
  \caption{Schematic picture of the capillary filling in a cylindrical tube with one end closed.}
  \label{fig1}
\end{figure}

Let us consider a tube of circular cross-section with radius $a$ and length $H$.
One end ($x=H$) of the tube is closed and the other end ($x=0$) is open.
The tube is initially filled with gas, and we immerse the open end into a fluid horizontally at time $t=0$ (Fig.~\ref{fig1}).
The fluid then starts to fill the tube, and we will discuss the filling dynamics $h(t)$.
In the horizontal geometry, we can neglect the effect of gravity.
We consider only the situation when inertial is not important, thus the fluid flow obeys Stokesian hydrodynamics.

When the fluid is filling the tube, there are three forces acting on the fluid.
\begin{itemize}
\item Capillary force
\begin{equation}
  F_{\rm c} = 2 \pi a \gamma_{\rm w} = 2 \pi a (\gamma_{\rm SV} - \gamma_{\rm SL}) = 2\pi a \gamma \cos \theta_E,
\end{equation}
where $\gamma_{\rm w} = \gamma_{\rm SV} - \gamma_{\rm SL}$ is the wicking parameter \cite{Quere2008}.
Using Young's relation we can write the force in terms of the equilibrium contact angle $\theta_E$ and fluid surface tension $\gamma$.
The capillary force is the driving force for fluid filling.

\item Viscous force
\begin{equation}
  F_{\rm v} = 8 \pi \eta h \dot{h},
\end{equation}
where $\eta$ is the fluid's viscosity, and $\dot{h}=\ud h / \ud t$ is the filling speed.
This is the frictional force against the filling.
\change{Here we have neglected the small contribution due to the meniscus because the meniscus has a much smaller volume in comparison to the bulk.}

\item Pressure force
\begin{equation}
  F_{\rm p} = (p_{\rm in} - p_0) \pi a^2,
\end{equation}
where $p_{\rm in}$ is the pressure of the trapped gas, and $p_0$ is the atmosphere pressure.
The pressure force can drive or prevent the filling, depending on the sign of $(p_{\rm in}-p_0)$.
\end{itemize}

In the following, we shall use the force balance of these forces to derive the dynamic equation for the filling length $h(t)$.

\subsection{Open model of Lucas-Washburn}
\label{sec:open}

We start with the classical model of Lucas and Washburn \cite{Lucas1918, Washburn1921}, where both ends of the tube are open.
In this case, the pressure force $F_{\rm p}$ vanishes because $p_{\rm in}=p_{\rm 0}$.
The force balance gives
\begin{equation}
  \label{eq:open1}
  2 \pi a \gamma_{\rm w} = 8 \pi \eta h \dot{h} \quad \Rightarrow \quad
  \dot{h} = \frac{a \gamma_{\rm w}}{4 \eta h} .
\end{equation}
The solution to Eq.~(\ref{eq:open1}) is
\begin{equation}
  \label{eq:LWh}
  h^2 = \frac{ a \gamma_{\rm w} }{2\eta} t.
\end{equation}
This is the classical Washburn-Lucas result with $h \sim \sqrt{t}$.

The equation (\ref{eq:open1}) can be written in a dimensionless form,
\begin{equation}
  \frac{\ud \bar{h}}{\ud \bar{t}} = \frac{1}{\bar{h}},
  \label{eq:open2}
\end{equation}
using the following transformation
\begin{equation}
  h = \bar{h} H, \quad t = \bar{t} \, \tau_{\rm LW} = \bar{t} \, \frac{4 \eta H^2}{a \gamma_{\rm w}}.
  \label{eq:open3}
\end{equation}
We have scaled the length in term of tube length $H$, and the time in the unit of $\tau_{\rm LW} = 4 \eta H^2 / (a \gamma_{\rm w})$.

\subsection{Insoluble gas model}

We now consider the closed-end case.
If the gas is insoluble in the fluid, the total number of trapped gas molecules remains constant.
When the filling continues, the enclosed gas is compressed and its pressure increases. Assuming ideal gas, the gas pressure obeys Boyle's law
\begin{equation}
  p_{\rm in} (H-h) = p_0 H.
\end{equation}
The pressure force $F_{\rm p}$ is written as
\begin{equation}
  F_{\rm p}=\frac{h}{H-h} p_0 \pi a^2.
\end{equation}
Balance of all three forces gives
\begin{eqnarray}
  2\pi a \gamma_{\rm w} &=& 8\pi \eta h \dot{h} + \frac{h}{H-h} p_0 \pi a^2 \\
  \dot{h} &=& \frac{a \gamma_{\rm w}}{4 \eta h} \Big[ 1 - \frac{h}{H-h} \frac{a p_0}{2 \gamma_{\rm w}} \Big]
  \label{eq:insoluble1}
\end{eqnarray}
Here a dimensionless number $\alpha = ap_0/2\gamma_{\rm w}$ appears, which characterizes relative magnitude of the pressure force and capillary force.

The dimensionless form of Eq.~(\ref{eq:insoluble1}) is
\begin{equation}
  \frac{\ud \bar{h}}{\ud \bar{t}} = \frac{1}{\bar{h}} \Big[ 1 - \frac{\bar{h}}{1-\bar{h}} \alpha \big].
\label{eq:insoluble}
\end{equation}
The above equation has an analytic solution \cite{Phan2010}
\begin{equation}
  \bar{t} = \frac{\bar{h}^2}{2(1+\alpha)} - \frac{ \alpha \bar{h}}{(1+\alpha)^2} - \frac{\alpha^3}{(1+\alpha)^3} \ln \big[ 1 - (1+\alpha)\bar{h} \big].
\end{equation}
The asymptotic solution is
\begin{equation}
  \label{eq:tinf}
  \bar{h}|_{\bar{t} \rightarrow \infty} = \frac{1}{1+\alpha}.
\end{equation}
The final length depends on the dimensionless number $\alpha$.

\subsection{Soluble gas model}
\label{subsec:real}

We now take one step further and consider the soluble gas.
Henry's law gives the relation between the number density of dissolved gas at the gas/fluid boundary and the gas pressure
\begin{equation}
  \label{eq:Henry}
  n = \frac{p}{k_H} ,
\end{equation}
where $k_H$ is the Henry's constant \cite{Sander2015}.
The filling process compresses the trapped gas, increases the pressure near the closed end, and the dissolved gas density at $x=h$ is greater than that at the open end $x=0$.
In this case, the dissolved gas will diffuse from the closed end to the open end.
The dissolved gas density obeys the diffusion-convection equation
\begin{equation}
  \label{eq:dcA}
  \frac{\partial n}{\partial t} = D_g \frac{\partial^2 n}{\partial x^2} - \dot{h} \frac{\partial n}{\partial x},
\end{equation}
where $n(x)$ is the number density of the dissolved gas.
The first term in Eq.~(\ref{eq:dcA}) comes from diffusion and $D_g$ is the diffusion constant of the dissolved gas in the fluid.
The second term is the convection term due to the fluid flow.
The boundary conditions to Eq.~(\ref{eq:dcA}) are given by the Henry's law
\begin{equation}
  n(0)=\frac{p_0}{k_H}, \quad n(h)=\frac{p_{\rm in}}{k_H},
  \label{eq:bcA}
\end{equation}
where $p_0$ and $p_{\rm in}$ are pressures at the open end and closed end, respectively.

The equation of state for trapped gas is given by
\begin{equation}
  \label{eq:ideal_gas}
  p_{\rm in} \pi a^2 (H-h) = N k_BT
\end{equation}
where $N$ is the number of trapped gas molecules (here we have assumed only one type of gas molecule).
\change{
At the gas/fluid interface near the closed end, the gas is dissolved at the rate
\begin{equation}
  \label{eq:dNdtA}
  \frac{\ud N}{\ud t} = - D_g \frac{\partial n}{\partial x} \Big|_{x=h} \pi a^2.
\end{equation}
Note there is no convection term here due to the moving boundary.}

The dynamics of the filling length $h(t)$ is given by the force balance
\begin{equation}
  \label{eq:dhdtA}
  2\pi a \gamma_{\rm w} = 8\pi \eta h \dot{h} + ( p_{\rm in} - p_0 ) \pi a^2.
\end{equation}

To summarize, the filling process is described by the dissolved gas density $n(x)$, the number $N$ of trapped gas molecules, and the filling length $h$.
The time evolution of $n(x)$, $N$, $h$ are given by Eqs. (\ref{eq:dcA}) [with Eq.~(\ref{eq:bcA}) as boundary conditions], (\ref{eq:dNdtA}), and (\ref{eq:dhdtA}), respectively.

We perform the dimensionless transformation.
Beside variable changes in Eq.~(\ref{eq:open3}), we also need the following relations
\begin{eqnarray}
  x = \bar{x} \, H , &\quad&
  n = \bar{n} \, \frac{p_0}{k_{\rm H}}, \\
  p = \bar{p} \, p_0, &\quad&
  N = \bar{N} \, \frac{p_0 H \pi a^2}{k_BT}.
\end{eqnarray}
The dimensionless form of the diffusion-convection equation (\ref{eq:dcA}) is
\begin{equation}
  \label{eq:dcA2}
  \frac{\ud \bar{n}}{\ud \bar{t}} = D \frac{\partial^2 \bar{n}}{\partial \bar{x}^2} - \frac{\ud \bar{h}}{\ud \bar{t}} \frac{\partial \bar{n}}{\partial \bar{x}},
\end{equation}
where the dimensionless diffusion constant $D$ is given by
\begin{equation}
  D = D_g \frac{4\eta}{a \gamma_{\rm w}}.
\end{equation}

The diffusion-convection equation in this case is a moving-boundary problem and numerically difficult.
We get around by the following change of variables
\begin{equation}
  z = \frac{\bar{x}}{\bar{h}}, \quad \tau = \bar{t}.
\end{equation}
The diffusion-convection equation (\ref{eq:dcA2}) becomes
\begin{equation}
  \label{eq:dcA3}
  \frac{\partial \bar{n}}{\partial \tau} = \frac{D}{\bar{h}^2} \frac{ \partial^2 \bar{n}}{\partial z^2} + \frac{z-1}{\bar{h}} \frac{\ud \bar{h}}{\ud \tau} \frac{\partial \bar{n}}{\partial z}.
\end{equation}
The boundary conditions to Eq.~(\ref{eq:dcA3}) are
\begin{equation}
  \bar{n} |_{z=0} = 1, \quad \bar{n} |_{z=1} = \bar{p}_{\rm in}.
\end{equation}
\change{We now work in the fixed boundaries, and the diffusion-convection equation (\ref{eq:dcA3}) is solved numerically using Crank-Nicolson method.}

The dimensionless forms of Eqs. (\ref{eq:dNdtA}) and (\ref{eq:dhdtA}) are
\begin{eqnarray}
  \frac{\ud \bar{N}}{\ud \tau} &=& - \frac{\kappa D }{\bar{h}} \frac{\partial \bar{n}}{\partial z} \Big|_{z=1},\\
  \frac{\ud \bar{h}}{\ud \tau} &=& \frac{1}{\bar{h}} \big[ 1 - \alpha (\bar{p}_{\rm in}-1) \big].
\end{eqnarray}
Here one more dimensionless number $\kappa = k_BT/k_{\rm H}$ is introduced.
The pressure of trapped gas is given by
\begin{equation}
  \label{eq:barPin}
  \bar{p}_{\rm in} = \bar{N}/(1-\bar{h}).
\end{equation}

\subsection{Linear profile models}
\label{sec:soluble_linear}

In Ref.~\cite{Phan2010}, Phan et al. had developed a similar model to consider the effect of dissolved gas.
Instead of solving a diffusion-convection equation (\ref{eq:dcA}), the authors assumed the dissolved gas density along the tube takes a linear profile
\begin{equation}
  \label{eq:linear}
  n(x) = n(0) + \frac{x}{h} \big( n(h)-n(0) \big).
\end{equation}
They also neglected the contribution from the convection term.
If the constant $D \gg 1$, the diffusion is a fast process and the linear assumption is valid.
When $D\sim 1$ or $D<1$, we need to solve the diffusion-convection equation numerically \cite{Weijs2013a, Lv2014Metastable}.
Here we shall present the model with linear profile for comparison.

\change{Since the density profile takes the linear form (\ref{eq:linear}), Eq.~(\ref{eq:dNdtA}) becomes
\begin{equation}
  \frac{\ud N}{\ud t} = - D_g \frac{ n(h)-n(0) }{h} \pi a^2.
\end{equation}}
The coupled dynamical equations in the dimensionless form (without convective term) can be derived as
\begin{eqnarray}
  \frac{\ud \bar{h}}{\ud \tau} &=& \frac{1}{\bar{h}} \big[ 1 - \alpha ( \bar{p}_{\rm in} - 1 ) \big], \\
  \label{eq:dNdtl}
  \frac{\ud \bar{N}}{\ud \tau} &=& \kappa D \frac{1-\bar{p}_{\rm in}}{\bar{h}}.
\end{eqnarray}

\section{Results and Discussions}
\label{sec:result}

We have introduced three dimensionless numbers to characterize the system,
\begin{eqnarray}
  \alpha &=& \frac{ap_0}{2 \gamma_{\rm w}} , \\
  D &=& \frac{D_g 4\eta}{a \gamma_{\rm w}} , \\
  \kappa &=& \frac{k_B T}{k_H} .
\end{eqnarray}
The number $\alpha$ is the ratio between the atmosphere pressure $p_0$ and the capillary pressure $2 \gamma_{\rm w} /a$.
This is the driving force for capillary filling.
The other two numbers are related to the dissolution of trapped gas molecules.
The escape of trapped gas molecules has two steps.
The first is the dissolution at the air/fluid interface, which is characterized by the dimensionless number $\kappa$.
The second step is the diffusion process from the closed end to the open end, which is characterized by the dimensionless diffusion constant $D$.

In this section, we present numerical results based on different models presented in Section 2.
We shall consider the time evolution of the filling length $\bar{h}$, the number of the trapped gas molecules $\bar{N}$, and the density profile of dissolved gas $\bar{n}(z)$ for various sets of parameters $(\alpha, D, \kappa)$.

In Table~\ref{tab:1}, we list physical constants and material properties of common fluids.
The values of three dimensionless parameters then can be calculated as below
\begin{eqnarray}
  \alpha = 0.00791, \quad D=0.6327, \quad \kappa = 0.0414.
\end{eqnarray}
In the subsequent sections, we use the approximated values as nominal values to perform computation
\begin{eqnarray}
  \alpha = 0.1, \quad D=1, \quad \kappa = 0.1.
\end{eqnarray}
We shall systematically examine the effect of each dimensionless number by varying one while keeping other two the same.

\vspace{0.2cm}
\begin{table}[htbp]
\centering
\begin{tabular}{lcrl}
  \hline\hline
  parameter  & symbol  & value & unit \\
  \hline
  gas diffusion constant   & $D_g$      & $1.0 \times 10^{-9}$ & m$^2$/s \\
  Henry constant           & $k_H$      & $1.0 \times 10^{-19}$  & Pa$\cdot$m$^3$ \\
  Boltzmann constant times temperature   & $k_B T$   & $1.38 \times 10^{-23} \times 300$ & J  \\
  fluid viscosity (water)  & $\eta$     & 0.1 & Pa$\cdot$s  \\
  surface tension (water)  & $\gamma$   & $7.3 \times 10^{-2}$ &  N/m \\
  atmosphere pressure      & $p_0$      & $1.0 \times 10^5$ & Pa  \\
  tube radius              & $a$        & $1.0 \times 10^{-8}$ &  m \\
  contact angle            & $\theta$   & $30$ & degree \\
  \hline\hline
\end{tabular}
\caption{List of physical constant and material properties}
\label{tab:1}
\end{table}

\subsection{Effect of ${\alpha}$}
\label{subsec:alpha}

\begin{figure}[htbp]
  \centering
  \includegraphics[width=1.0\columnwidth]{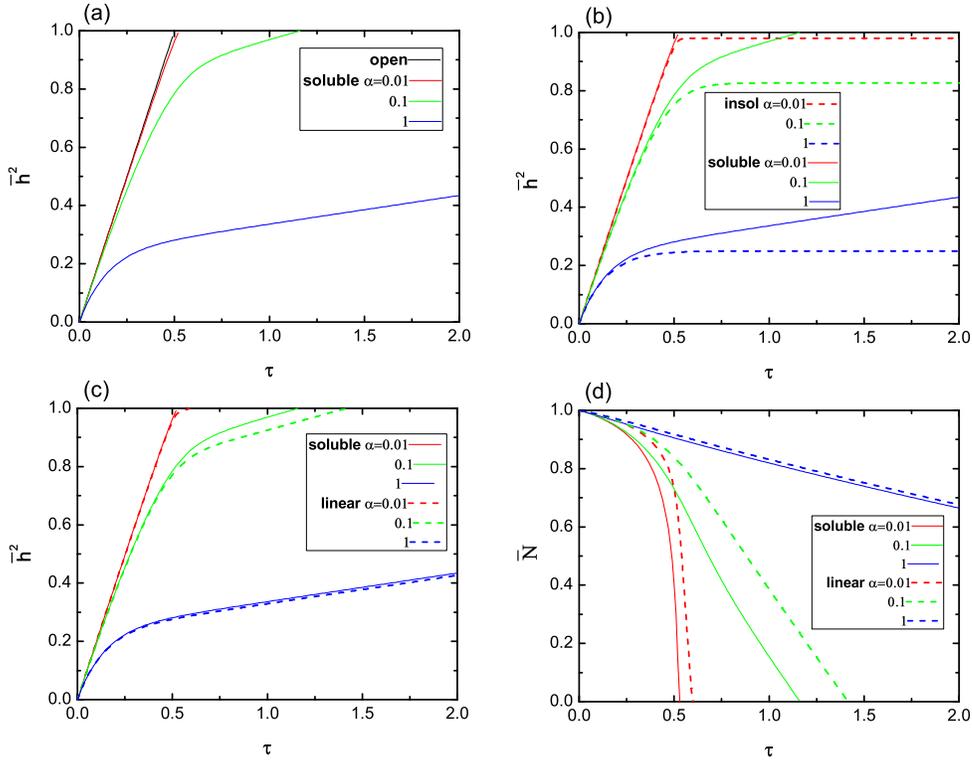}
  \caption{\change{The time evolution of the filling length $\bar{h}^2(\tau)$ [(a)--(c)] and the number of trapped gas $\bar{N}(\tau)$ [(d)]. The value of $\alpha$ is varied while the other two are set as $D=1$ and $\kappa=0.1$. (a) Comparison between open model and soluble gas model. (b) Comparison between insoluble gas model and soluble gas model. (c) and (d) Comparison of soluble gas model and linear model.}}
  \label{fig:alpha1}
\end{figure}

The dimensionless number $\alpha$ is the ratio between atmosphere pressure and the capillary pressure.
The increase of $\alpha$ corresponds to the increase of the tube radius $a$ or the decrease of the wicking parameter $\gamma_{\rm w}$.
Figure \ref{fig:alpha1} shows the effect of varying $\alpha$ for systems with $D=1$ and  $\kappa=0.1$.

The effect of varying $\alpha$ on the time evolution of the filling length $\bar{h}$ is shown in Fig.~\ref{fig:alpha1}(a), where we compare the results for the open model and soluble gas model.
The open model follows the Lucas-Washburn dynamics: a straight line of $\bar{h}^2 \sim \tau$.
For soluble gas model, the dynamics can be separated into two stages: At the initial time, the filling is fast and the process resembles the open model.
At later time, the trapped gas is compressed and the pressure contribution becomes important, resulting a slowdown in the filling dynamics.
Large $\alpha$ corresponds to a small capillary pressure and a weak driving force, therefore the slowdown is more pronounce for large $\alpha$ values.

In Fig.~\ref{fig:alpha1}(b), we compare the results for the insoluble gas model and soluble gas model.
The insoluble model predicts a plateau of the filling length at later time.
The plateau is the result of balance between the capillary pressure and the gas pressure difference between the closed and open ends.
The asymptotic length decreases when $\alpha$ increases, consistent with prediction of Eq.~(\ref{eq:tinf}).

\change{The comparisons for soluble gas model and linear model are shown in Fig.~\ref{fig:alpha1}(c) and \ref{fig:alpha1}(d).
For $D=1$ and $\kappa=0.1$, the linear model gives similar evolution for $\bar{h}$ as the soluble gas model, but the dynamics is slower.
The slow-down is more pronounced at later stage when the evolution starts to deviate from the linear Lucas-Washburn behavior. 
The difference is more obvious at intermediate $\alpha$ value.
This can be seen more clearly in the plot of $\bar{N}$ [Fig.~\ref{fig:alpha1}(d)].
At $\alpha=1$, the curves for $\bar{N}$ are similar, while for $\alpha=0.1$, the number of trapped gas show different evolution for the two models. 
For small $\alpha$ value, the decrease of trapped gas at later time is very quick, so the difference in $\bar{h}$ only appears in a short period after $\tau=0.5$.}

\begin{figure}[htbp]
  \centering
  \includegraphics[width=1.0\columnwidth]{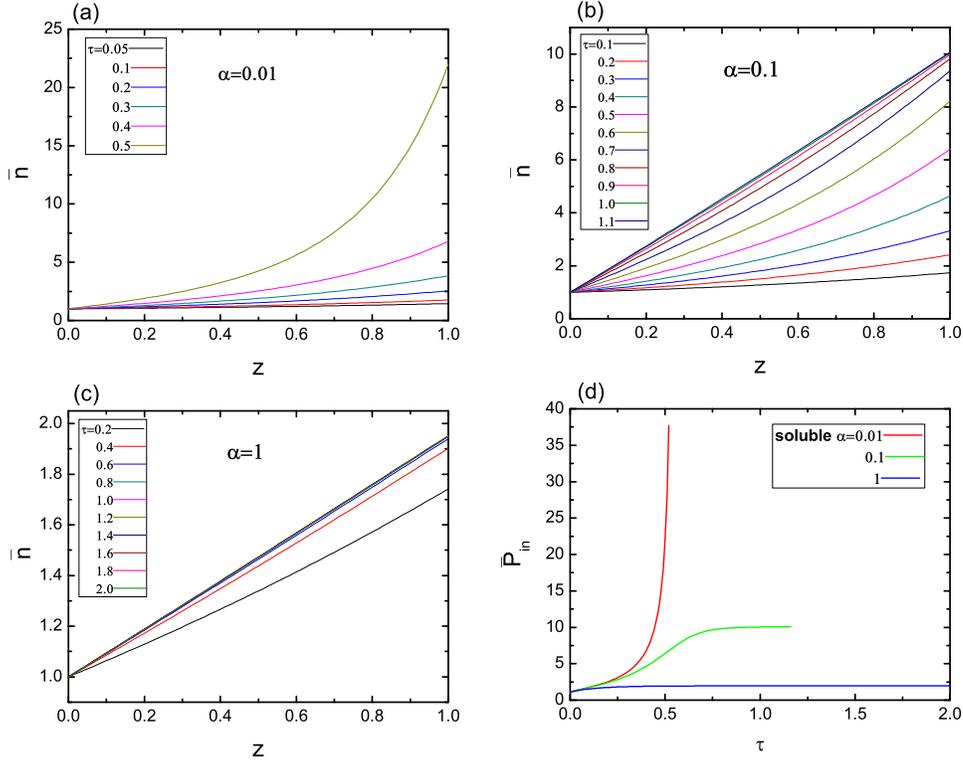}
  \caption{\change{The evolution of the density profile of dissolved gas $\bar{n}(z,\tau)$ [(a)--(c)] and the pressure of trapped gas $\bar{p}_{\rm in}(\tau)$ for the soluble gas model [(d)].  The value of $\alpha$ is varied while the other two are set as $D=1$ and $\kappa=0.1$.}}
  \label{fig:alpha2}
\end{figure}

In Fig.~\ref{fig:alpha2}(a)--(c), we show the density profiles at different time for the soluble gas model.
The density profiles are convex at initial times and becomes more linear at later times.
The linear profiles are only present for large $\alpha$ values; when $\alpha=0.01$, the fluid has already reached the tube end before the linear profile appears.
The development of nonlinear profiles is associated with a rapid increase of the pressure in the trapped gas [Fig.~\ref{fig:alpha2}(d)].
Once the pressure reaches its plateau value and changes slowly, the density profiles become linear.

\subsection{Effect of $D$}

\begin{figure}[htbp]
  \centering
  \includegraphics[width=1.0\columnwidth]{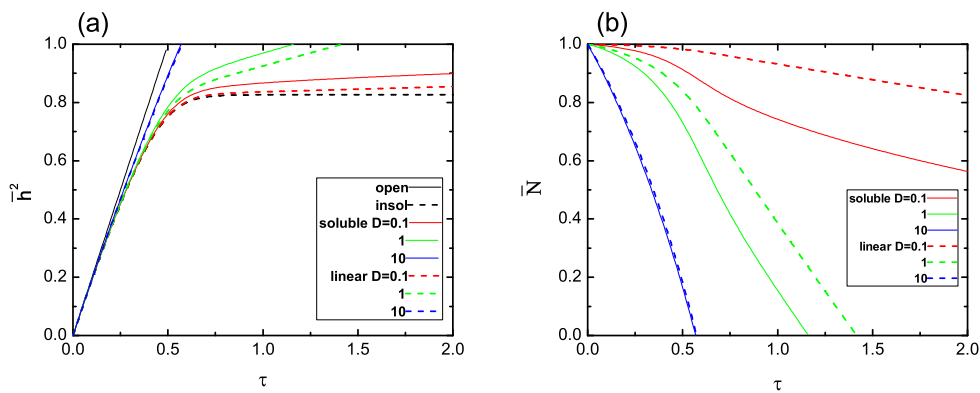}
  \caption{\change{The time evolution of the filling length $\bar{h}^2(\tau)$ [(a)] and the number of the trapped gas $\bar{N}(\tau)$ [(b)]. The value of $D$ is varied while the other two are set as $\alpha=0.1$ and $\kappa=0.1$.}}
  \label{fig:D1}
\end{figure}

We next examine the effect of diffusion constant $D$.
The dimensionless parameter $D$ represents how fast the dissolved gas moves from high density region to low density region.
For large value of $D$, any density variation is quickly flattened by the diffusion, and one would expect that the linear approximation works well.
Figure \ref{fig:D1} shows the time evolution of $\bar{h}^2$ and $\bar{N}$ for different models.
The parameters are $\alpha=0.1$ and $\kappa = 0.1$.

From Fig.\ref{fig:D1}(a), when $D$ is a large, the evolutionary trends of the filling length for different models are close to that of open model, while when $D$ is small, the evolution approaches to the insoluble gas model.
From Fig.\ref{fig:D1}(a) and \ref{fig:D1}(b), we observe that when $D=10$, the evolution for different models differs little.
As $D$ decreases, the difference appears and is enhanced when $D$ becomes smaller.
This observation agrees with our reasoning in Section 2.3:
When the constant $D \gg 1$, the diffusion is a fast process and the linear assumption can be applied, therefore the linear models give reasonable results.
When $D\sim 1$ or $D<1$, the situation is quite different and we need to make precise calculation.

\begin{figure}[htbp]
  \centering
  \includegraphics[width=1.0\columnwidth]{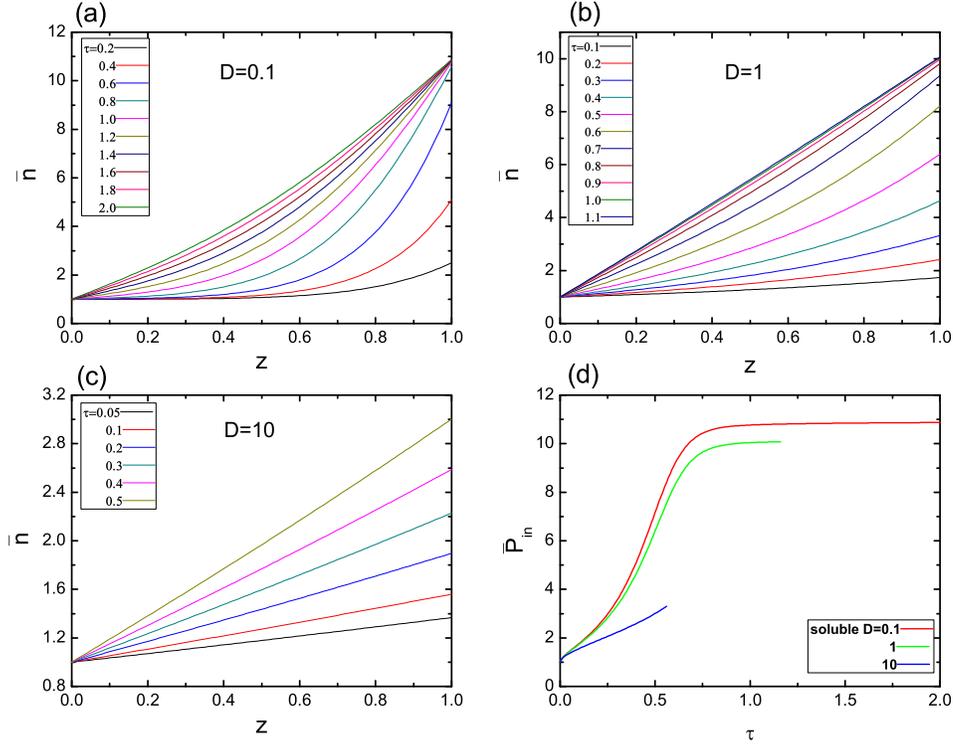}
  \caption{\change{The evolution of the density profile of dissolved gas $\bar{n}(z,\tau)$ [(a)--(c)] and the pressure of trapped gas $\bar{p}_{\rm in}(\tau)$ for the soluble gas model [(d)].  The value of $D$ is varied while the other two are set as $\alpha=0.1$ and $\kappa=0.1$.}}
  \label{fig:D2}
\end{figure}

In Figure~\ref{fig:D2}, we show the evolution of the density profiles and the pressure of trapped gas.
The profiles are nonlinear for $D=0.1$, and becomes more linear when $D$ increases.
This is consistent with the results of filling length [Fig.~\ref{fig:D1}(a)]: linear models differ from the soluble gas model only when $D$ is small.

\subsection{Effect of $\kappa$}

\begin{figure}[htbp]
  \centering
  \includegraphics[width=1.0\columnwidth]{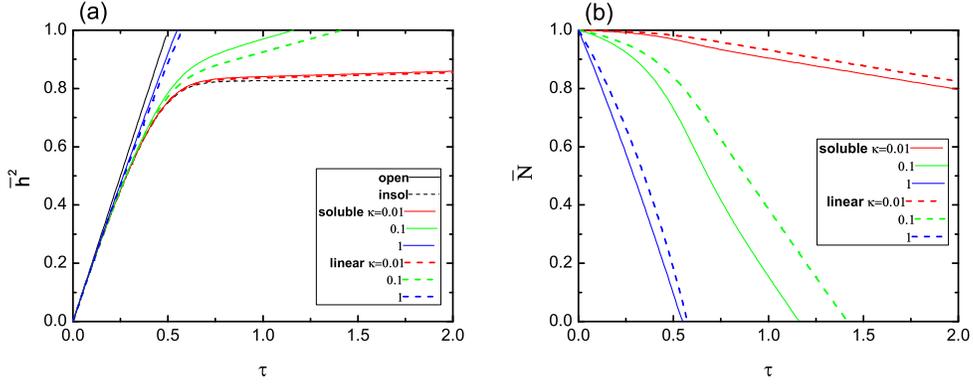}
  \caption{\change{The time evolution of the filling length $\bar{h}^2(\tau)$ [(a)] and the number of the trapped gas $\bar{N}(\tau)$ [(b)]. The value of $\kappa$ is varied while the other two are set as $\alpha=0.1$ and $D=1$.}}
  \label{fig:B1}
\end{figure}

From the expression of $\kappa$, we see that the change of this parameter will cause different dissolving rate in the interface between the trapped gas and fluid.
The effect of $\kappa$ on the filling dynamics is shown in Fig.~\ref{fig:B1} for $\alpha=0.1$ and $D=1$.

From Fig.~\ref{fig:B1}(a), one can see when $\kappa$ is small, the evolution resembles insoluble gas model, while when $\kappa$ is large, the filling becomes fast and resembles the open model.
This is similar to the results of Section 3.2.
It is clearly that different $\kappa$ values lead to different evolutionary trends.

\begin{figure}[htbp]
  \centering
  \includegraphics[width=1.0\columnwidth]{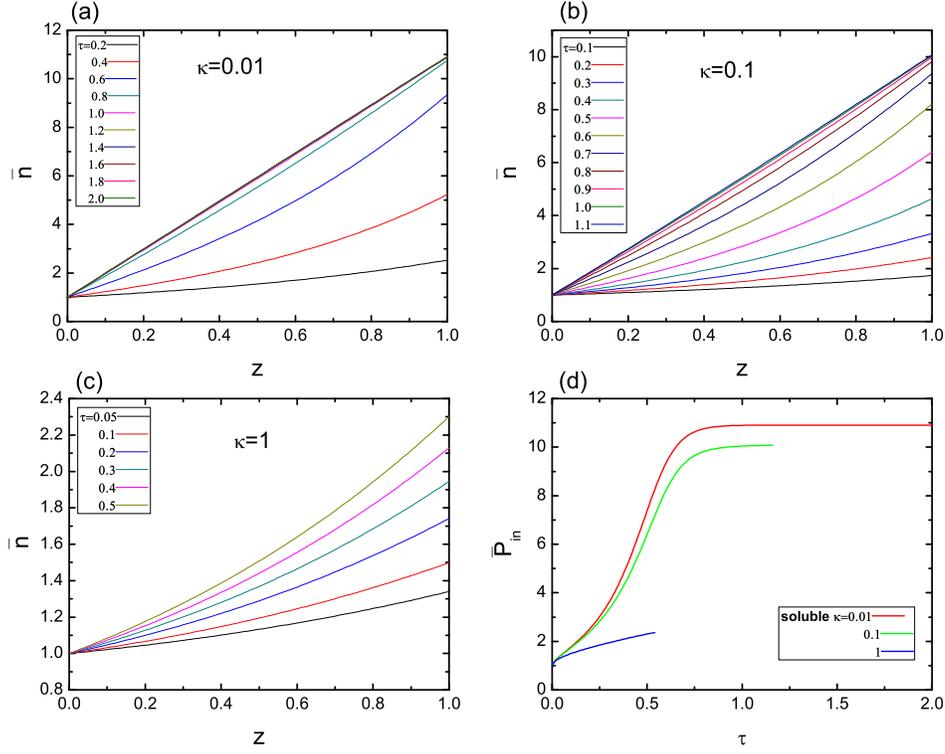}
  \caption{\change{The evolution of the density profile of dissolved gas $\bar{n}(z,\tau)$ [(a)--(c)] and the pressure of trapped gas $\bar{p}_{\rm in}(\tau)$ for the soluble gas model [(d)].  The value of $\kappa$ is varied while the other two are set as $\alpha=0.1$ and $D=1$.}}
  \label{fig:B2}
\end{figure}

The evolution of the density profiles and the pressure of trapped gas for different $\kappa$ values are shown in Fig.~\ref{fig:B2}.
The profiles exhibit similar trends as shown in Fig.~\ref{fig:alpha2} and Fig.~\ref{fig:D2}.
The slope of the profile changes from large to small when $\kappa$ increases (note the scale changes in the $y$-axis).

\subsection{Filling time}

\begin{figure}[htbp]
  \centering
  \includegraphics[width=0.8\columnwidth]{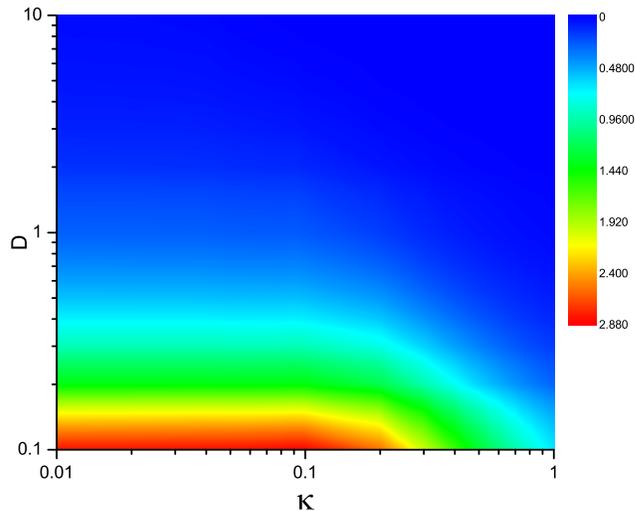}
  \caption{\change{The time difference [Eq.~(\ref{eq:deltaT})] between the soluble gas model and the linear model. The difference is shown in the plane of $\kappa$--$D$. The red region corresponds to large time difference.}}
  \label{fig:time2}
\end{figure}

To show the difference between different soluble gas models, we compute the dimensionless time $\bar{T}$ for the filling length to reach $\bar{h}=0.99$.
In Fig.~\ref{fig:time2}, we plot the time difference between the soluble gas model and linear model in the parameter space of $\kappa$--$D$,
\begin{equation}
\label{eq:deltaT}
\Delta \bar{T} = \bar{T}_{\rm linear} - \bar{T}_{\rm soluble}.
\end{equation}

When $D$ is large, one can safely use the linear approximation of the density profile, because the time difference $\Delta \bar{T}$ is very small.
When $D$ and $\kappa$ are both small, the time difference becomes large, thus one has to resort to the soluble gas model and solve the diffusion-convection equation.

\section{Summary}
\label{sec:summary}

In this paper, we present a model on the capillary filling in closed-end nanotubes.
We made two improvement on previous model of Ref.~\cite{Phan2010}:
\begin{enumerate}
\item[(1)] We do not make the assumption that the density profile of dissolved gas is linear.
\item[(2)] We explicitly include the convection term in the model.
\end{enumerate}

The filling dynamics is characterized by three dimensionless numbers ($\alpha$, $D$, $\kappa$) and we systematically examine the time evolution for different parameter sets.
Our results show that when the gas dissolution and diffusion is slow, one need to solve the diffusion-convection equation rigorously.
Assumption of linear profile and neglect of convection term may not be appropriate in these situations.

\begin{acknowledgments}
This project was supported by the National Natural Science Foundation of China (Grant No.~21434001, 21504004, and 21774004).
\end{acknowledgments}

\bibliography{closed_tube}

\end{document}